\documentstyle[aps,prl,twocolumn,epsf,floats,12pt]{revtex}

\draft
\begin{document}
\title{Reversed-Spin Quasiparticles 
       in Fractional Quantum Hall Systems 
       and Their Effect on Photoluminescence}
\author{
   Izabela Szlufarska$^{1,2}$, 
   \underline{Arkadiusz W\'ojs}$^{1,2}$, and 
   John J. Quinn$^1$}
\address{\footnotesize\sl
   $^1$Department of Physics, 
       University of Tennessee, Knoxville, Tennessee 37996, USA \\
   $^2$Institute of Physics, 
       Wroclaw University of Technology, Wroclaw 50-370, Poland\\[1em]}
\address{
   \footnotesize\rm\parbox{6.0in}{
   The energy, interaction, and optical properties of reversed-spin
   quasielectrons (QE$_{\rm R}$'s) in fractional quantum Hall systems 
   are studied.
   Based on the short range of the QE$_{\rm R}$--QE$_{\rm R}$ repulsion, 
   a partially unpolarized incompressible $\nu={4\over11}$ state is 
   postulated within Haldane hierarchy scheme.
   To describe photoluminescence, a reversed-spin fractionally charged 
   exciton $h$QE$_{\rm R}$ (QE$_{\rm R}$ bound to a valence hole $h$) 
   is predicted.
   In contrast to its spin-polarized analog, $h$QE$_{\rm R}$ is strongly 
   bound and radiative.
   }
   \\[-3ex]}
\maketitle

Laughlin quasielectrons (QE's) and quasiholes (QH's) of fractional 
quantum Hall (FQH) systems\cite{laughlin,prange} can be thought of 
as empty (QH) or filled (QE) states of the lowest or the first excited 
composite fermion (CF) Landau level (LL)\cite{jain}, respectively.
The energies of these Laughlin quasiparticles (QP's) and their 
interactions with one another and with a valence hole ($h$) have 
been studied quite thoroughly by means of exact diagonalization of 
small systems.
The influence of QP's on photoluminescence (PL) of the Laughlin 
electron fluid is important in those systems in which the hole 
is spatially separated from the electrons by a distance $d$ that 
exceeds approximately one magnetic length ($\lambda$). 
In such systems, PL occurs from the radiative bound states of a hole 
and one or two QE's, called fractionally charged excitons (FCX's)
\cite{wojs-fcx}.
Reversed-spin QE's (denoted by QE$_{\rm R}$'s)\cite{rezayi,chakraborty} 
are another type of elementary excitations of a Laughlin fluid, which 
can be thought of as particles in the reversed-spin lowest CF LL.
As other QP's, the QE$_{\rm R}$'s carry fractional charge and have finite 
size, energy, and angular momentum.

In this note, the single-particle properties of the QE$_{\rm R}$'s, 
as well as the pseudopotentials defining their interaction with one 
another and with other QP's, are determined numerically and used to 
predict when the incompressible-fluid states with less than maximum 
polarization occur.
The interaction of QE$_{\rm R}$'s with valence holes is also 
studied as a function of the layer separation $d$.
In analogy to the FCX states, stable reversed-spin FCX's (denoted 
by FCX$_{\rm R}$) are predicted at a finite $d$ and small Zeeman 
splitting.
It is shown that the FCX$_{\rm R}$ optical selection rules are 
different from those of FCX's, because of the QE$_{\rm R}$ angular 
momentum and spin being different from those of a QE.
For example, the ground state of one QE$_{\rm R}$ bound to a hole 
is optically active, in contrast to the nonradiative $h$--QE pair 
ground state.
The stability of the radiative FCX and FCX$_{\rm R}$ states, and 
thus also the PL spectrum of the Laughlin fluid, is shown to depend 
on the layer separation $d$, Zeeman splitting, and (critically) on 
the electron filling factor $\nu$.

In order to preserve the 2D translational symmetry of infinite systems,
in a finite-size calculation we use Haldane spherical geometry
\cite{haldane} in which the LL degeneracy $g=2S+1$ is controlled by 
the strength $2S$ of the magnetic monopole placed in the center of 
the sphere of radius $R$.
The monopole strength $2S$ is defined in the units of flux quantum
$\phi_0=hc/e$, so that $4\pi R^2B=2S\phi_0$ and $R^2=S\lambda^2$.
The many-body states on the sphere are labeled by the set of good 
quantum numbers: electron spin ($J$) and its projection ($J_z$), 
hole spin ($\sigma_h$), and the length ($L$) and projection ($L_z$) 
of the total angular momentum of the electron-hole system.

The ground state of the 2DEG in the lowest LL at the Laughlin filling 
factor $\nu={1\over3}$ is completely polarized in the absence of the 
Zeeman splitting, $E_{\rm Z}=0$.
There are two types of elementary charge-neutral excitation of these 
ground states, carrying spin $\Sigma=0$ or 1, which in literature are 
referred to as the charge- and spin-density waves, respectively.
The most important features of their dispersion curves 
${\cal E}_{\Sigma}(k)$ ($k$ is the wave vector) are the magneto-roton 
minimum at the finite value of $k=1.5\,\lambda^{-1}$ in ${\cal E}_0(k)$, 
the finite gap $\Delta_0\approx0.076\,e^2/\lambda$ at this minimum, 
and the vanishing of ${\cal E}_1$ in the $k\rightarrow0$ limit.
Equally important is the similarity of the charge- and spin-density 
waves in the $\nu={1\over3}$ state to those at $\nu=1$.
The latter can be understood by means of Jain CF picture\cite{jain} 
where the excitations of the $\nu=1/3$ electron state correspond 
to promoting one CF from a completely filled lowest $(n=0)$ 
spin-$\downarrow$ CF LL either to the first excited ($n=1$) CF LL 
of the same spin ($\downarrow$) or to the same CF LL ($n=0$) but 
with the reversed spin ($\uparrow$).

One can define three types of QP's (elementary excitations) of the 
Laughlin $\nu={1\over3}$ fluid that constitute the charge- and 
spin-waves: Laughlin QH's and QE's and Rezayi QE$_{\rm R}$'s.
Each of the QP's is characterized by such single-particle quantities 
as electric charge (${\cal Q}_{\rm QH}=+{1\over3}e$ 
and ${\cal Q}_{\rm QE}={\cal Q}_{\rm QER}=-{1\over3}e$), degeneracy 
$g_{\rm QP}$ of the single-particle Hilbert space, and energy 
$\varepsilon_{\rm QP}$.
The charge-neutral excitations of Laughlin ground states are composed 
of a pair of QH and either QE ($\Sigma=0$) or QE$_{\rm R}$ ($\Sigma=1$).

In order to estimate the energies $\varepsilon_{\rm QP}$ needed 
to create an isolated QP of each type, we applied the exact 
diagonalization procedure to systems of $N\le11$ electrons.
By extrapolating the results to $N\rightarrow\infty$, we found the 
following values appropriate for an infinite system: 
$\varepsilon_{\rm QE}=0.0664\,e^2/\lambda$ and 
$\varepsilon_{\rm QER}=0.0383\,e^2/\lambda$. 
Our estimation of so-called ``proper'' QP energies (obtained by 
adding the term ${\cal Q}_{\rm QP}^2/2R$ to $\varepsilon_{\rm QP}$) 
are: $\tilde{\varepsilon}_{\rm QE}=0.0737\,e^2/\lambda$,
$\tilde{\varepsilon}_{\rm QER}=0.0457\,e^2/\lambda$, and 
$\tilde{\varepsilon}_{\rm QH}=0.0258\,e^2/\lambda$.
Consequently, the energies of spatially separated QE--QH and 
QE$_{\rm R}$--QH pairs are equal
${\cal E}_0=\tilde{\varepsilon}_{\rm QE}+
\tilde{\varepsilon}_{\rm QH}=0.0995\,e^2/\lambda$ and 
${\cal E}_1=\tilde{\varepsilon}_{\rm QER}+
\tilde{\varepsilon}_{\rm QH}=0.0715\,e^2/\lambda$
(these are activation energies in transport experiments).

Which of the two negatively charged QP's (QE or QE$_{\rm R}$) occur 
at low energy in the particular system depends on the Zeeman term 
which is influenced not only by magnetic field, but also by many 
material parameters.
Once the QP's content is established, the correlations in the 
system can be understood by studying the appropriate interaction 
pseudopotentials defined as the dependence of pair interaction 
energy $V$ on the relative pair angular momentum ${\cal R}$ 
(larger ${\cal R}$ corresponds to larger separation)
\cite{wojs-hierarchy}.
Here, ${\cal R}_{\rm QE-QER}=l_{\rm QE}+l_{\rm QER}-L$ and
${\cal R}_{\rm QE-QE}=l_{\rm QE}+l_{\rm QE}-L$,
where $l_{\rm QP}$ and $L$ denote the one- and two-QP angular 
momenta, respectively.
The QH--QH, QE--QE, QE--QH, and QE$_{\rm R}$--QH pseudopotentials can 
be found elsewhere\cite{wojs-hierarchy,iza}, and therefore we will 
limit this discussion to $V_{\rm QER-QER}$ and $V_{\rm QE-QER}$ only.

\begin{figure}[t]
\epsfxsize=3.20in
\epsffile{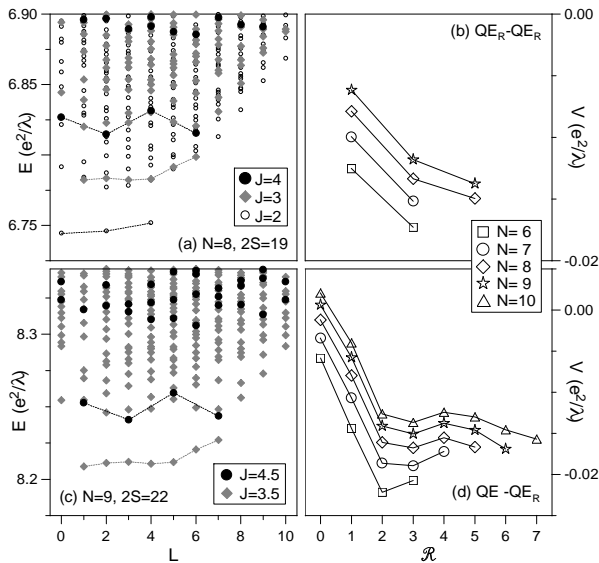}
\caption{(ac) The $N$-electron energy spectra calculated on Haldane
sphere at different values of $2S$.
The pseudopotentials of the QE$_{\rm R}$--QE$_{\rm R}$ (b) and
QE--QE$_{\rm R}$ (d) interactions calculated for $N\le10$.}
\label{fig1}
\end{figure}

Two QE$_{\rm R}$'s can be formed in a $N$-electron system with 
at least two reversed spins at $2S=3(N-1)-2$.
An example of such spectrum is shown in Fig.~\ref{fig1}(a) for $N=8$ 
with $J=2$, 3, and 4 corresponding to two, one and zero reversed 
spins, respectively.
It is clear that the maximally spin-polarized system ($J={1\over2}N$) 
is unstable at filling factors not equal to the Laughlin value 
$\nu={1\over3}$.

The $V_{\rm QER-QER}$ is shown in Fig.~\ref{fig1}(b).
Although the obtained values depend on the system size (the repulsive 
character of interaction is restored only in the $N\rightarrow\infty$ 
limit with $V_{\rm QER-QER}(1)\approx0.01\,e^2/\lambda$), the 
monotonicity of $V_{\rm QER-QER}$ seems to be independent of $N$.
Moreover, the super-linear shape of the curve indicates Laughlin 
correlations and thus incompressibility at $\nu_{\rm QER}={1\over3}$,
${1\over5}$, \dots (in analogy to Haldane hierarchy picture of completely 
spin-polarized states\cite{haldane,wojs-hierarchy}).
For example, Laughlin $\nu_{\rm QER}={1\over3}$ state occurs at the
electron filling factor of $\nu={4\over11}$ and corresponds to the
75\% spin polarization (this state has also recently been proposed 
in the CF model\cite{park}).
In view of the fact that the spin-polarized $\nu_{\rm QE}={1\over3}$ 
state is compressible\cite{wojs-hierarchy}, the experimental observation
\cite{stormer} of the FQH effect at $\nu={4\over11}$ confirms the 
formation of QE$_{\rm R}$'s in the $\nu={1\over3}$ state.

An QE--QE$_{\rm R}$ pair can be formed in the system with at least 
one reversed spin.
Another example of such spectrum is shown in Fig.~\ref{fig1}(c) for 
$N=9$ at $2S=3(N-1)-2=22$.
The lowest energy states in the two considered subspaces, 
$J={1\over2}N={9\over2}$ and $J={1\over2}N-1={7\over2}$, contain 
a QE--QE and QE--QE$_{\rm R}$ pair, respectively.
The pseudopotential $V_{\rm QE-QER}({\cal R})$ was calculated for 
$N\le10$ and the results are presented in Fig.~\ref{fig1}(d).
The values of the pseudopotential depend on $N$ and in the limit 
$N\rightarrow\infty$ we found V$_{\rm QE-QER}(0)\rightarrow0.015\,
e^2/\lambda$ and $V_{\rm QE-QER}(1)\rightarrow0.01\,e^2/\lambda$.

The behavior of $V_{\rm QE-QER}({\cal R})$ is qualitatively different 
from that of $V_{\rm QER-QER}({\cal R})$ and  $V_{\rm QE-QE}({\cal R})$.
The most significant feature of the $V_{\rm QE-QER}({\cal R})$ function
is that it is super-linear in $L(L+1)$ only at $1\le{\cal R}\le3$ and 
sub-linear at $0\le{\cal R}\le2$ and at larger ${\cal R}$.
As a result, $m=2$ is the only possible Jastrow exponent in the 
many-body wave function that describes Laughlin correlations and 
thus yields incompressibility of the system.

If QE's and QE$_{\rm R}$'s could coexist in the $\nu={1\over3}$ 
``parent'' state (unlikely due to their sensitivity to Zeeman 
energy), one could apply a generalized CF picture\cite{wojs-cf} 
to predict the allowed combinations of Jastrow exponents 
[$m_{\rm QE-QE},m_{\rm QER-QER},m_{\rm QE-QER}$] describing the 
incompressible states of such two-component plasma.
Based on the behavior of the three involved QP pseudopotentials 
for different values of ${\cal R}$, we reduced the number of 
possible exponent combinations to a few, from which only [1,1,2] 
satisfies the incompressibility condition.
Such hypothetical mixed QE/QE$_{\rm R}$ state corresponds to the 
$\nu={5\over13}$ state with 80\% spin polarization.

The PL spectra of a spin-polarized 2DEG can be understood in terms 
of QE's and their interaction with one another and with a valence 
hole ($h$), where electron and hole layers are separated by a finite 
distance $d$ (of the order of $\lambda$).
It was shown\cite{wojs-fcx} that in the regime where the 
electron-electron repulsion is weak compared to the electron-hole 
attraction, a hole can bind one or two QE's to create FCX ($h$QE 
or $h$QE$_2$).
The optical selection rules following from the 2D translational 
symmetry leave $h$QE$^*$ (the asterisk denotes an excited state) 
and $h$QE$_2$ to be the only optically active (``bright'') states.
By similar consideration, in the partially unpolarized systems we 
expect the positively-charged $h$ to bind one or two QE$_{\rm R}$ 
forming states denoted here by FCX$_{\rm R}$.
However, the radiative recombination of FCX$_{\rm R}$ will be governed 
by different selection rules due to different angular momenta of QE and
QE$_{\rm R}$ and thus different angular momenta of initial (bound) states
($l_{\rm FCXR}\ne l_{\rm FCX}$).

Because of strong QE$_{\rm R}$--QE$_{\rm R}$ repulsion, binding of more 
than one QE$_{\rm R}$ by the valence hole $h$ should be more difficult.
Similarly to the QE case, the simplest FCX$_{\rm R}$, $h$QE$_{\rm R}$, 
is expected to occur in a system containing free QE$_{\rm R}$'s at the 
values of $d$ at which the binding energies of $h$QE and $h$QE$_{\rm R}$ 
are smaller than the Laughlin gap to create additional QE-QH pairs.
We have studied a good number of numerical spectra for different values 
of $d$.
As expected, we found that both $h$QE$_{\rm R}$ and $h$QE ground states 
develop at $d$ larger than $\lambda$.

\begin{figure}[t]
\epsfxsize=3.20in
\epsffile{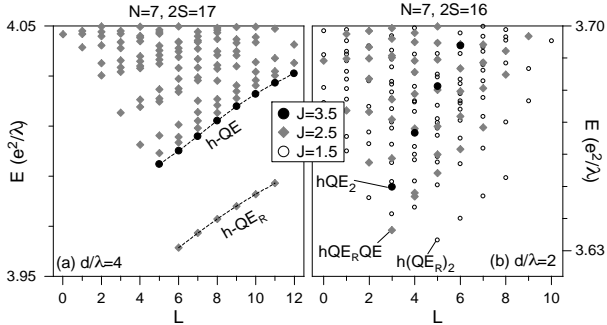}
\caption{The energy spectra of a system of seven electrons and
one hole on Haldane sphere at $d/\lambda=4$ with $2S=17$(a) 
and at $d/\lambda=2$ with $2S=16$ (b).}
\label{fig2}
\end{figure}

An example of such spectrum for a 7$e$--$h$ system (with up to one 
reversed electron spin) at $d=4\lambda$ and $2S=17$ is shown in 
Fig.~\ref{fig2}(a).
In the CF picture this configuration corresponds to six CF's filling 
the first CF LL and the seventh CF lying either in the second CF LL 
with parallel spin ($J={1\over2}N={7\over2}$) or in the first CF LL 
with reversed spin ($J={1\over2}N-1={5\over2}$).
As shown in Fig.~\ref{fig2}(a), at sufficiently large $d$ the lowest 
energy states contain well defined $h$--QE or $h$--QE$_{\rm R}$
pairs with all possible values of $L$ resulting from the rules of 
addition of angular momenta.
Applying the appropriate orbital selection rule for radiative
recombination\cite{wojs-fcx,iza} we determined that, unlike the dark 
$h$QE, the reversed-spin $h$QE$_{\rm R}$ ground state is radiative.
As shown in Fig.~\ref{fig2}(b) for $d=2\lambda$, the numerical 
spectra contain also larger FCX$_{\rm R}$ complexes, 
$h$(QE$_{\rm R}$)$_2$ and $h$QE$_{\rm R}$QE.
However, due to their weaker binding and because $h$(QE$_{\rm R}$)$_2$
turns out dark and $h$QE$_{\rm R}$QE is very sensitive to Zeeman energy,
the emission of $h$QE$_{\rm R}$ is expected to dominate the PL spectrum
of a partially unpolarized system at $\nu\approx{1\over3}$.

The authors acknowledge support of grant DE-FG02-97ER45657 
from Materials Science Program -- Basic Energy Sciences of the 
US Dept.\ of Energy.
IS and AW acknowledge support of Polish KBN grant 2P03B11118.

\end{document}